\documentclass[aps,prb,twocolumn,floatfix,showpacs]{revtex4}
\usepackage{epsfig,amsmath,amssymb,color}

\begin{document}

\title{Designing a symmetry protected molecular device}
\author{C. A. B\"usser}
\affiliation{Department of Physics and Astronomy, University of Wyoming, Laramie, Wyoming 82071, USA}
\author{A. E. Feiguin}
\affiliation{Department of Physics and Astronomy, University of Wyoming, Laramie, Wyoming 82071, USA}

%% The \maketitle command is necessary to build the title page.

\date{\today}
\begin{abstract}
Realizing a quantum transistor built of molecules or quantum dots has been one of the most ambitious challenges in nanotechnology. Even though remarkable progress has been made, being able to gate and control nanometer scale objects, as well to interconnect them to achieve scalability remains extremely difficult. Most experiments concern a single quantum dot or molecule, and they are made at ultra low temperature to avoid decoherence and tunneling. We propose to use canonical transformations to design quantum devices that are protected by symmetry, and therefore, may be operational at high temperatures. We illustrate the idea with examples of quantum transistor architectures that can be connected both in series and parallel. 
\end{abstract}


\maketitle

\section{Introduction} Thanks to advances in nanofabrication, experimentalist
can routinely manufacture nanostructures that resemble artificial atoms
--quantum dots--
that can be manipulated with an extreme degree of control\cite{quantum-dots1,quantum-dots2}. The degree of tunability is such, that one can see a state with a single electron sitting at the dot, and even control and detect its spin. By tuning a gate voltage one can also control the transport through the dot in what is regarded as a single-electron transistor. The techniques used to construct these devices can also be applied to build structures with multiple dots, and suggest that scalability to circuits of many quantum dots should be technologically achievable.
Another avenue that has been pursued is using molecules to build transistors, instead of quantum dots\cite{molecules}. Some examples of working devices have been realized with nanotubes\cite{cnt_transistor1,cnt_transistor2}, benzene molecules\cite{benzene_transistor,benzene_transistor2,benzene3,Xu2005}, fullerene\cite{c60_transistor,kondo_c60}, graphene \cite{graphene_qd,graphene_transistor}, and proteins\cite{Chen2012}, to mention a few. Adding contacts and gating these devices is very challenging, and scalability may be more difficult to attain. 
A third possibility consists of manipulating single atoms to build complex structures \cite{Fiete2003} that could be controlled using an STM tip\cite{Smit2003}.

Whether a complex molecule, or a quantum-dot array, the most general transport features of these systems in the absence of interactions can infered from some basic fundamental physics: Aharonov-Bohm-type interference\cite{Aharonov1959}, and Fano resonances\cite{Fano1961}. There is an important effort toward engineering devices at the atomic/molecular scale taking advantage of these different phenomena, and that can have useful functionality as classical switches --diod, transistor--, or novel behavior that could open the doors to quantum circuits with no classical analogy. These systems could lead to a new generation of electronic devices, and revolutionary applications such as quantum computers.

The chemistry of these molecules, or the geometry of the circuits, can be simple, or not. In this work we show that regardless of the chemistry, the transport properties of a molecular device can be inferred from symmetry considerations\cite{benzene1,DiVentra2000}. Moreover, canonical transformations allow us to establish dualities between quantum circuits, providing us with a tool to design devices that are protected by symmetry. 

%For simplicity we are going to start by considering the case of the ``circuit'' depicted in Fig.1. It is a simple chain of atoms with hopping $t$, with a couple of tunnelling barriers  rbitrary tunnelling amplitudes $t1$ ane $t2$. In addition, there is a chemical potential $\mu$ on the site labelled as ``b''. We now proceed to  

\begin{figure}
\epsfig {file=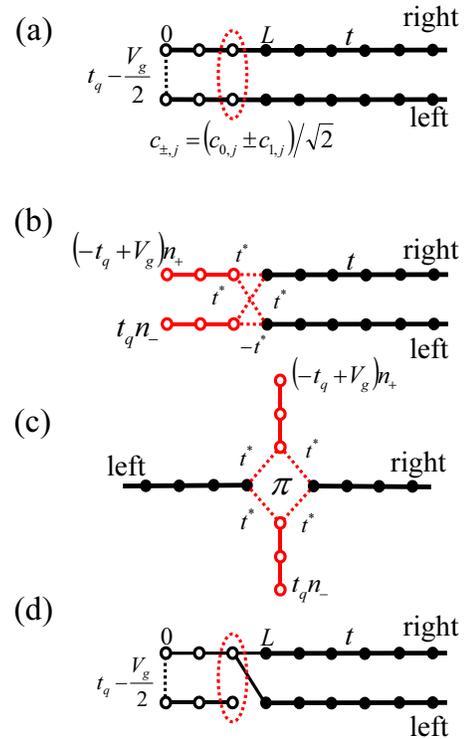,width=80mm,angle=0}
\caption{
Illustration showing the original system (a), and the equivalent one after the folding transformation is applied to the sites $0,...L-1$ (b) and (c). The emerging $\pi$ phase can be eliminated by starting from the circuit (d).
 The value of the hopping connecting the two halves becomes $t^*=t/\sqrt{2}$ after the transformation.} \label{map1}
\end{figure}

\section{Equivalent quantum circuits}

Consider a one dimensional chain of atoms, described by a tight-binding chain with hopping $t$, with a weak link in the center, with hopping $t_q$. Let us split the chain into left and right halves, and fold it in two, as depicted in Fig. \ref{map1}(a). The corresponding Hamiltonian is:
\begin{eqnarray}
H & = & H_\mathrm{leads} + H_\mathrm{link} \nonumber \\
H_\mathrm{leads} & = & -t\sum_{\lambda=0,1}\sum_{j=0}\left(c_{\lambda,j}^\dagger c_{\lambda,j+1} + \mathrm{h.c.}\right) \\
H_\mathrm{link} & = & (-t_q+\frac{V_g}{2}) \left(c_{0,0}^\dagger c_{1,0} + \mathrm{h.c.}\right) + \frac{V_g}{2} \left(n_{0,0}+n_{1,0}\right) \nonumber
\label{hami}
\end{eqnarray}
where $c_{\lambda,j}$ is the electron annihilation operator acting on site $j$ of lead $\lambda$ (where the values 0,1 correspond to left and right leads), and sites labeled ``0'' connect the two leads. The addition of the term proportional to $V_g$ will become apparent below. For the moment, we take $V_g=0$, and consider spinless fermions, but these considerations can be easily generalized to the spin-full case.
We now introduce a symmetric (+) and antisymmetric (-) combination of operators acting on the left and right leads. This is nothing else but an application of the reflection symmetry, yielding new even(+) and odd(-) operators:
\begin{equation}
c_{\pm,j} = \frac{1}{\sqrt{2}}(c_{0,j} \pm c_{1,j}).
\end{equation}
This is a simple change of basis, with the new operators obeying the same fermionic anti-commutation rules. This is a well known canonical transformation, originally introduced in the context of quantum impurity problems \cite{Kane1992b}, and referred to as a ``folding'' transformation \cite{simon2001}. It is used extensively in Numerical Renormalization Group (NRG) calculations \cite{Bulla2008}, and it has been recently used to study impurity problems on rings using the density matrix renormalization group (DMRG) \cite{canonical}. It has also been used in rather creative ways to simplify complex problems in quantum chemistry, such as conjugated molecules and polymers \cite{Ramasesha1992}. 
In this work, we shall take an unconventional approach: instead of applying the transformation to the entire system, we shall apply it to only a fraction of it, say $L$ sites to the right of the weak link. As a result, we find a section of the chain transformed to the new basis, connected to the rest of the left and right leads that remain in the original basis, as shown in Fig. 1(b). At the boundary connecting the transformed and untransformed sections, we find new hopping terms mixing the two bases as:
\begin{eqnarray}
H_{\mathrm{boundary}} & = & -\frac{t}{\sqrt{2}} \left[ c_{0,L}^\dagger \left(c_{+,L-1} + c_{-,L-1} \right)\right. \nonumber \\
& + & \left. c_{1,L}^\dagger \left(c_{+,L-1} - c_{-,L-1} \right) \right] + \mathrm{h.c.}
\label{boundary}
\end{eqnarray}
After some simple algebra, it is easy to see that the hopping at the weak link, maps onto effective chemical potentials in the new basis (see Appendices A and B for details):
\begin{equation}
H_\mathrm{link} = -t_q \left(n_{+,0}-n_{-,0}\right) + V_g n_{+,0},
\label{link}
\end{equation}
where $n_\pm=c^\dagger_\pm c_\pm$ is the density operator. This term is just a boundary chemical potential with opposite signs for the ``+'' and ``-'' sections of the leads, at site ``0''. All the other hopping terms along the ``+'' and ``-'' sections of the lead remain unaltered in the new basis.

\begin{figure}
\epsfig {file=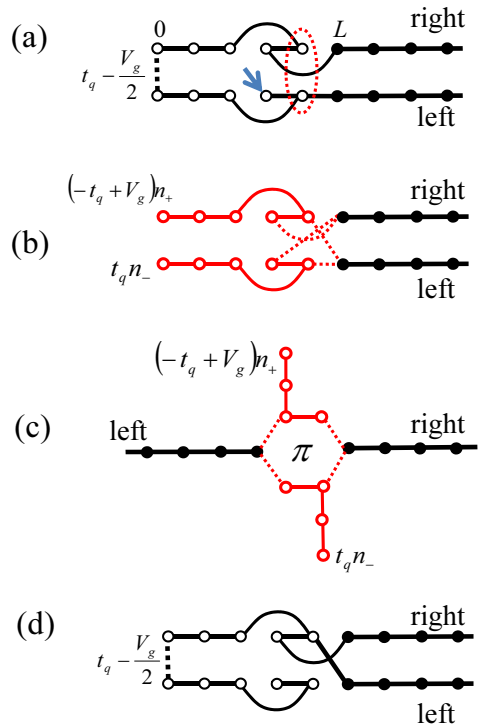,width=80mm}
\caption{
Illustration showing the the transformation that leads to an annulene-like geometry: (a) original system,  and the equivalent one after the folding transformation is applied to the sites $0,...L-1$ (b) and (c). The emerging $\pi$ phase can be eliminated by starting from the circuit (d).
The arrow in panel (a) indicates a dangling link.
} \label{map2}
\end{figure}

Taking all this into account, and after some visual inspection, it is easy to see that the original chain, after the basis transformation, maps onto the equivalent circuit shown in Fig. \ref{map1}(c): a diamond connected to the left and right leads (that remain in the original basis), and the ``+'' and ``-'' leads that have a finite length, and a boundary chemical potential at the ends of magnitude $t_q$. Let us call these leads ``gates'' from now on.
All hoppings inside the diamond are equal to $t/\sqrt{2}$, with the exception of one link that has the opposite sign. This can be interpreted as a $\pi$ magnetic flux threading the diamond.

The physics of the original system, and the new equivalent one should be completely equivalent, since this mapping is exact. Therefore, let us consider setting the boundary chemical potential of the two ``+'' and ``-'' leads on the equivalent circuit to zero, $t_q=0$. If we look at the original chain, this corresponds to effectively cutting the connection between left and right leads, completely opening the circuit. Therefore, under these conditions, no current can circulate from left to right leads, no matter what the temperature is. The circuit is completely open, by simple symmetry considerations! 

If one wanted to describe the transport properties of the equivalent circuit, one could use ideas of quantum interference\cite{Sautet1988,Cardamone2006,Rincon2009,Bulgakov2006}, and would consider the anti-resonances coming from the gates\cite{Orellana2003b} or combinations of the two \cite{Jana2008}. However, this simple mapping shows a different aspect of the problem, and allows one to understand all the transport properties by simply studying at the original unfolded tight-binding chain.

\begin{figure}
\epsfig{file=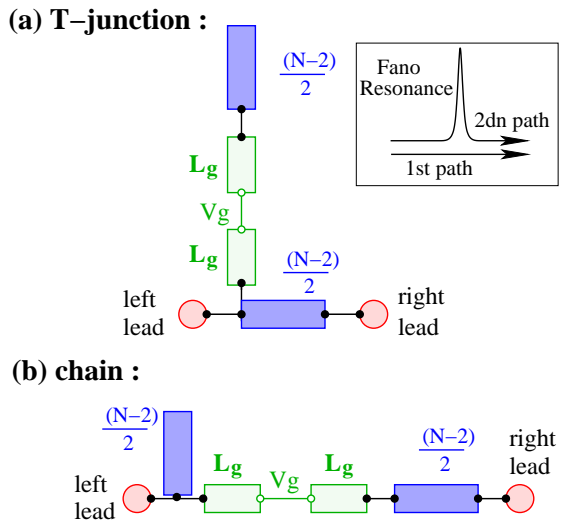,width=8.5cm}
\caption{Rearrangement of the atoms in the circuit: (a) single tight-binding chain and (b) T-junction. In both cases a portion of the system is hanging, leading to an interference effect due to a Fano resonance, as depicted in the inset.}
\label{scheme}
\end{figure} 

So far, we have shown that the equivalent circuit acts like a multi-gate transistor: by simply tuning the potential at the gates, we can open the circuit, and completely suppress transport between the leads. Alternatively, at finite tunneling amplitude $tq \neq 0$, we can obtain perfect transport. Another situation that we can consider is connecting one of the gate leads to ground, and the other one to a gate voltage $V_g$. This corresponds to choosing $t_q=0$, $V_g \neq 0$ in Eq.(\ref{link}):
\begin{equation}
H_\mathrm{link} = V_g n_{+,0}.
\label{link2}
\end{equation}
%It can be easily seen that this corresponds to adding a chemical potential in the original unfolded basis, but preserving the geometry of the problem. 
%Therefore, we can study the transport properties using the tigh-binding chain, as a function of $V_g$. 
The results for the conductance are shown in Fig.\ref{Cuad-Oct}: At zero voltage $V_g=0$, the circuit is open, while at large voltages we recover the perfect conductance. This behavior is analyzed below, and in Appendix C.

\section{Canceling the magnetic flux}

One of the undesired aspects of the previous ``device'' is the emergence of a $\pi$ flux threading the loop. One could easily devise a setup to get rid of the phase. Consider the circuit shown in Fig.\ref{map1}(d), corresponding to a T-shape junction. After applying the folding transformation to all the sites situated to the left of the connecting link, we recover the same molecular circuit with the diamond in the center, but with no phase shift. The price we have to pay is that we can no longer interpret the problem in terms of a simple tight-binding chain. However, this still is a non-interacting circuit, and we can easily understand the transport properties. In this case, the conductance will depend on the parity of the dangling lead in the T-junction. Let us first consider the $V_g=0$ case. If $t_q \neq 0$ this segment will have an even number of sites, with a vanishing density of states at $\omega=0$. This means that it cannot participate in the transport between the left and right leads, yielding a perfect conductance. If $t_q = 0$, the segment is effectively cut in half, and now two situations can arise: (i) for $L=4k$, we obtain the same result as before for $t_q \neq 0$, or (ii) for $L=4k+2$, a resonance will appear at the Fermi level, and the conductance will cancel identically. Notice that these are more complex generalizations of the problem studied in Ref.\cite{Kubala2002}. The general situation for finite $V_g$ will be studied below. 

\section{General case: Annulenes}

We can generalize the previous protocol to other geometries, to obtain the equivalent of molecular transistors with a ring structure of $N$ sites, similar to the proposed devices based on annulenes\cite{ring_transistor}. 
Consider the geometry illustrated in Fig.\ref{map2}(a). It depicts a one dimensional chain, that has been folded in half at the weak link, but with the addition of a second ``twist'' plus as dangling link, marked with an arrow in Fig.\ref{map2}(a). By rotating the $(N-1)/2$ operators to the left of the $L$th site, we obtain the hexagonal molecule shown in Fig.\ref{map2}(b), connected to leads, and gates, as in the original case described before, with the emerging $\pi$ flux. The main difference with the tight-binding chain is the dangling sites, which introduce an additional degree of complexity. If the size of the annulus is $N=4k$, the dangling link will have an even number of sites, and will not participate in the transport properties, yielding the same results as the diamond. For $N=4k+2$, the dangling link will introduce an anti-resonance, and the conductance will be $G=0$.

%Results for the conductance are shown in Figures\ref{Cuad-Oct} and \ref{Hex-Dec}. For $N=4k$, the behavior is similar to the original tigh-binding chain, while for $N=4k+2$, the conductance behavior is reversed, with perfect conductance at the particle-hole symmetric point $V_g=0$. In any case, the devices behave as a three-gate transistor, and are protected by symmetry.

Again, the undesired $\pi$ phase can be removed by re-shaping the circuit as a $T$-junction (see Fig.\ref{map2}(d)). In this case, we expect a parity effect in the transport behavior, depending on the size of the ring $N$, and the size of the dangling ``gate leads'', $L_g=L-(N-2)/2$.

\section{Transport properties}
%%%%%%%%% On the conductance calculations...
We now analyze the case with $t_q=0$, $V_g \neq 0$ in more detail.
In general, the transport properties are related to the Green's functions that propagate one electron from the left to the right leads $G_{LR}(\omega)$.
The conductance $G$ at zero bias can be written as \cite{wingreen}
\begin{equation}
G = \frac{e^2}{h} t^4 |G_{LR}(E_F)|^2 \rho_0(E_F)^2,
\end{equation}
where $\rho_0$ is the local density of states (LDOS) at the leads (assumed to be semi-infinite tight-binding chains) and $E_F=0$ is the Fermi level.

%To represent the leads we use half-filled semi-infinite tight binding chains with a local density of states at the first site of the chains given by
%\begin{equation}
%\rho_0(\omega) = \frac{\omega \pm \sqrt{\omega^2 - 4t^2}}{2t^2}.
%\end{equation}

Since the system is non-interacting, we can calculate the exact Green's functions using either the equations of motion formalism\cite{Zubarev} or by applying Wick's theorem. 

%%%%%%%%% TEXT FOR DISCUSSION OF THE FIGURES
In Figure~\ref{scheme} we have rearranged the geometry of the leads to show more clearly the structure of the system. Independently of the phase, the system always presents a dangling segment.
The conductance is controlled by Fano resonances. Two different paths, as represented in the inset of Fig.~\ref{scheme}(a), can interfere in a constructive or destructive way\cite{Orellana2003}.
The parity of the number of sites in the ring will determine whether the phases cancel or add up.

\begin{figure}
\epsfig{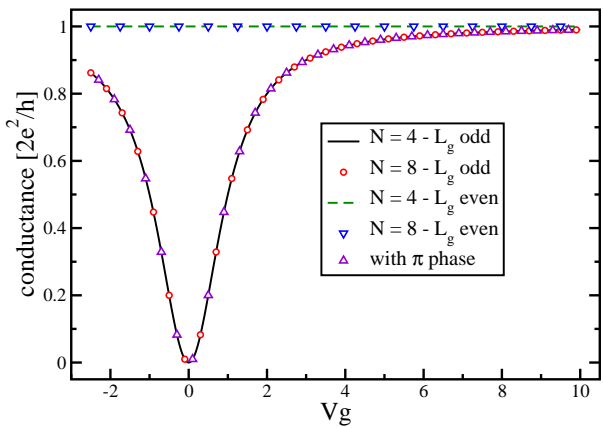}
\caption{Conductance for rings with 4 and 8 sites ($N=4k$ sites), and $t_q=0$. Two regimes appear: when $L_g$ is even, the conductance is perfect $2e^2/h$, while for $L_g$ odd, it strongly depends on $V_g$. }
\label{Cuad-Oct}
\end{figure}

%%%%%%%%% Text for Fig. Cuad-Oct.
It is convenient to start the discussion by considering the T-junction problem. In this case, the transport properties depend on both, the parity of $N$, and the size of the gate leads $L_g$. We first look at a ring with $N=4k$ sites. In Figure~\ref{Cuad-Oct} we present the results for the conductance for rings with 4 and 8 sites as a function of $V_g$ and gate leads with different lengths $L_g$, odd and even.
We find two clearly distinct situations depending on the parity of $L_g$. As shown in this figure, when $L_g$ is even a perfect conductance of $G=2e^2/h$ independent of $V_g$ is realized, while for odd $L_g$ the conductance depends strongly of the gate potential $V_g$. 
%%%%%%%%%%%%%%%%%%%%%%%%
To understand this behavior we look at Figure \ref{scheme}(a). As explained before, in the $V_g=0$ limit the upper part of the system is disconnected and the conductance is controlled by the number of sites in the gate leads $L_g$. 
If $L_g$ is odd, this segment will have a resonance at the Fermi level and again, this will cancel the conductance identically. 
If $L_g$ is even, there will not be any resonance in the LDOS and the conductance in this case is perfect, G=$2e^2/h$.

When $V_g \neq 0$ the upper part of the system of Figure \ref{scheme}(a) will be connected and the chemical potential at the ends will also start playing a role. The total number of sites of the hanging part will be $2L_g+(N-2)/2$ always odd, producing, in principle, a resonance at the Fermi level, and canceling the conductance. However, at large $V_g$ the chemical potential removes {\it one} site from the hanging segment, with the conductance growing toward $2e^2/h$ with increasing $V_g$. 

In order to understand the effects of a finite gate voltage $V_g$, let us consider the system before the transformation, as shown in Figures \ref{map1}(a) and (d), and take $L=1$, such that we apply the folding transformation only on the link connecting left and right leads $H_{\mathrm{link}}$ (\ref{link2}). This will yield two orbitals ``+'' and ``-'' that will be connected to the rest of the system through $H_{\mathrm{boundary}}$ (\ref{boundary}). In the large $V_g$ limit, the orbital ``+'' will be empty, and effectively removed from the system, which will have one site/orbital less. This changes the parity of the problem, converting a resonance into an anti-resonance and viceversa.

\begin{figure}
\epsfig{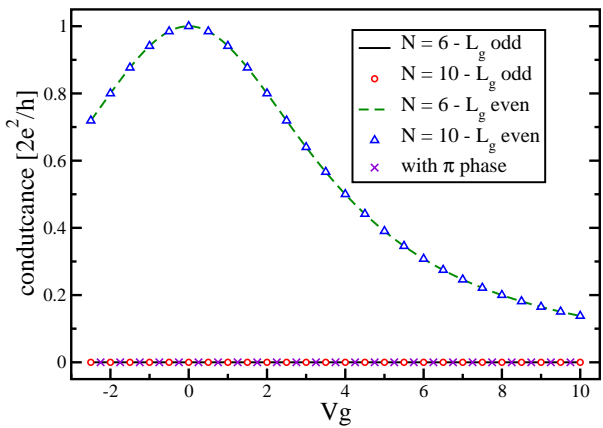}
\caption{Conductance for rings with 6 and 10 sites ($N=4k+2$ sites) and $t_q=0$. Again, two different regimes are realized: When $L_g$ is odd the conductance is always zero. For $L_g$ even, the conductance depends strongly on $V_g$.}
\label{Hex-Dec}
\end{figure}

%%%%%%%%% Text for Fig. Hex-Dec.
We now analyze rings with $N=4k+2$ sites. Results for the conductance as function of $V_g$ for rings with $6$ and $10$ sites are shown in Fig.~\ref{Hex-Dec}. 
Same as before, we can observe two different regimes. For $L_g$ odd the conductance is always zero while for $L_g$ even the conductance depends strongly on $V_g$.
The physics can be explained following similar arguments as the previous case, but with odd and even $L_g$ cases inverted.
%. The upper part of the hanging chain is uncoupled and the transport properties are controlled by the length $L$ of one block. For $L$ odd, the Fano resonance completely cancels the conductance, while for $L$ even the conductance will be perfect as shown in the figure. 
%
%For $V_g \neq 0$, the upper end of the hanging system is connected and a local chemical potential appears at the sites that connect the two blocks of size $L$. Now the total number of sites for the hanging part will be $2L+(2N-2)/2)$ and, as $N=2(2n+1)$, will be an even number. Then the LDOS of the hanging part will not have a resonance at the Fermi level. However the chemical potentials will push one resonance to the $\omega=0$ when $V_g$ increases, canceling the conductance.

%%%%%%%%%%%%%%%%%%%%%%%%%%%%%%%%%%%%%%%%%%%%%%%
We now go back to the case with a flux, corresponding to a single chain. As shown in Figures \ref{Cuad-Oct} and \ref{Hex-Dec}, the conductance is independent of $L_g$, as we should expect for a tight-binding chain. However, there is a parity effect corresponding to the size of the ring being $N=4k$ or $N=4k+2$, which dominates the transport properties. 
As mentioned before, this is explained by the effect of the dangling links added to the chain to construct the rings. Depending on the number of sites in the links, we will observe a resonance or anti-resonance at the Fermi level (see Fig.\ref{scheme}(b)). 
%For $N=4k+2$ this segment will have a resonance at the Fermi level $\omega=0$ in its LDOS. The hybridization of this resonance with the wire will cancel the conductance identically. 
%For $N=4k$ the segment has zero density of state at $\omega=0$, and it will not participate in the transport, yielding perfect conductance.

The effect of $V_g$ can again be explained as gradually removing a site from the chain (Appendix B). However, since adding or removing a site from a chain does not have any effect on its fundamental transport properties, the conductance interpolates smoothly between the $t_q=0,V_g=0$ and $t_q\neq 0,V_g=0$ cases.

For both $N=4k$, and $N=4k+2$ cases, with $L_g$ odd, the behavior is exactly the same for the chain and the T-junction. This is a remarkable result, and leads one to think of a hidden symmetry between the two setups. As a matter of fact, it is not a symmetry relating the two Hamiltonians, but a symmetry in the Green's functions that determine the equilibrium transport properties, that is ``magically'' realized when these quantities are evaluated at the Fermi level, $\omega=0$.  
This is analyzed in great detail in the Appendix C.

\section{Molecule in parallel}

Let us look again at Fig.\ref{map1}(d). We can make the transformation inside the hanging segment of the T-junction, away from the contact to the leads, as shown in Fig.\ref{map3}(a). It is easy to see that the equivalent circuit results as depicted in Fig.\ref{map3}(b). The physics of the problem, of course, is just as described in the previous sections. It is the same system after all! This shows that connecting the molecule in parallel, or in series, will have exactly the same effect and functionality. Moreover, it may be easier to realize experimentally, and more convenient for scalability purposes.

\section{Summary and Conclusions}

\begin{figure}
\epsfig {file=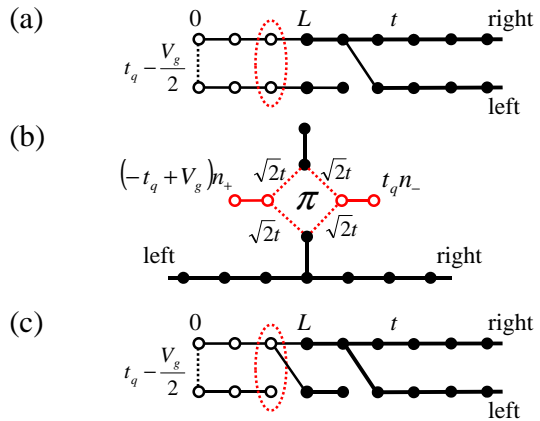,angle=-90,width=90mm}
\caption{Illustration showing the equivalence between (a) a T-junction, and (b) a molecular device connected in parallel. The equivalent circuit without a phase is depicted in (c).
} \label{map3}
\end{figure}

We have illustrated how three complex quantum ``devices'', shown in Figures \ref{map1}(c), \ref{map2}(c), and \ref{map3}(b) can be understood in terms of simple equivalent circuits that have the general topology of either a single chain, or a T-junction. These exact mappings have been obtained as a straightforward application of the folding transformation. We can easily understand the transport properties of these simple circuits, and explain the general behavior of the molecular device. Two remarkable aspects are that: (i) in the single chain case, with a phase $\pi$ threading the molecule, the gate voltage can be used to effectively ``cut'' the equivalent circuit in two, yielding a complete cancellation of the conductance. This effect is protected by the symmetry of the problem, and indicates that this behavior as a quantum switch survives at arbitrary temperatures. (ii) The zero-temperature transport properties of the chain and the T-junction are the same for $L_g$ odd, and can be understood in terms of a symmetry in the equilibrium Green's functions at $\omega=0$.

 Even though we have limited our analysis to a single orbital situation, these considerations can be generalized to a multi-orbital case. The non-interacting limit assumed here applies to a problem that can be described in terms of a H\"uckel-like theory. The addition of interactions would introduce long range terms in the transformed basis, making the folding transformation impractical. However, one could imagine studying the interactions within a Hartree-Fock framework, in which the resulting effective Hamiltonian is quadratic. In this picture, the interactions introduce a shift in the diagonal site energies, and the main effect is to shift and to split the resonance positions \cite{Chen1994,Yu1998}. In such scenario, as long as the reflection symmetry is preserved, our treatment is still valid.

Inteference-based molecular transistors have already been proposed \cite{Cardamone2006,Kocherzhenko2010,Kocherzhenko2011}, and quantum interference has recently been observed in molecules\cite{Guedon2012}. This study indicates a path toward understanding and designing molecular devices with customized functionality, and we illustrate it with setups that operate as multi-terminal transistors that can work connected in series, and parallel. These symmetries make us see these systems in a new light, and can also aid {\it ab-initio} calculations. It is clear that the mapping is not limited to the particular examples presented here, but it is more general and may lead to complex and interesting possibilities that could pave the way toward new advances in nano-electronics. 

\begin{acknowledgments}
AEF is grateful to NSF for funding under grant DMR-0955707.
\end{acknowledgments}

\section*{APPENDIX A: SYMMETRY TRANSFORMATIONS}

\begin{figure}
\epsfig {file=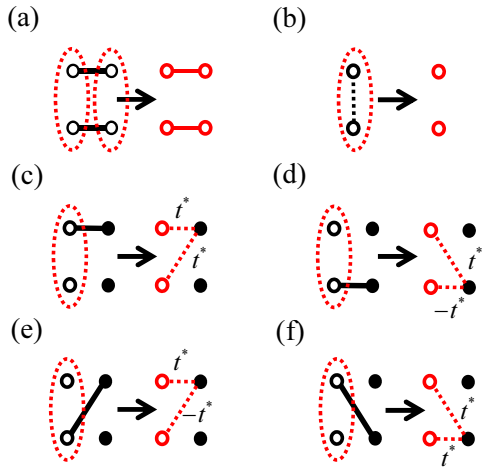,width=80mm,angle=0}
\caption{
Diagrams showing the resulting circuit after the transformation is applied to the encircled sites on the left. 
}
\label{appendix2}
\end{figure}

The entire treatment utilized in this work relies on a simple and basic symmetry : reflection, equivalent to a bonding/anti-bonding transformation.
\begin{equation}
c_{\pm,j} = \frac{1}{\sqrt{2}}(c_{0,j} \pm c_{1,j}) \,\,\, , \,\,\, c_{0/1,j} = \frac{1}{\sqrt{2}}(c_{+,j} \pm c_{-,j}).
\end{equation}

We start by considering a non-interacting one-dimensional chain, folded in two.
It is easy to understand the effects of this symmetry by visualizing them in diagrams, as depicted in Figure \ref{appendix2}. When this transformation is applied to two parallel links, as in diagram \ref{appendix2}(a), the transformation leaves the Hamiltonian invariant and reads exactly the same in the transformed basis:
\begin{eqnarray}
\sum_{\lambda=0,1}\left(c^\dagger_{\lambda,i}c_{\lambda,j}+\mathrm{h.c.}\right) & = & \sum_{\lambda=+,-}\left(c^\dagger_{\lambda,i}c_{\lambda,j}+\mathrm{h.c.}\right).
\end{eqnarray}

 When applied to a link connecting a site on chain ``0'' and its partner on chain ``1'', it leads to a vanishing hopping, and the introduction of two chemical potentials (see diagram \ref{appendix2}(b)):
\begin{eqnarray}
c^\dagger_{0,i}c_{1,i}+\mathrm{h.c.} & = & n_{+,i}-n_{-,i}, \\
n_{0,i}+n_{1,i} & = & n_{+,i} + n_{-,i}.
\end{eqnarray}

The interesting situation arises when the transformation is applied to the left of certain boundary, and not to the sites to the right. Then, at the link connecting the transformed sites, and those remaining in the original basis, new hopping terms emerge. The four possibilities shown in Figure \ref{appendix2}(c)-(f) correspond to:

\begin{eqnarray}
c^\dagger_{0,i}c_{0,j} & = & \frac{1}{\sqrt{2}} \left(c^\dagger_{0,i}c_{+,j}+c^\dagger_{0,i}c_{-,j}\right) \\
c^\dagger_{1,i}c_{1,j} & = & \frac{1}{\sqrt{2}} \left(c^\dagger_{0,i}c_{+,j}-c^\dagger_{0,i}c_{-,j}\right) \\
c^\dagger_{1,i}c_{0,j} & = & \frac{1}{\sqrt{2}} \left(c^\dagger_{1,i}c_{+,j}+c^\dagger_{1,i}c_{-,j}\right) \\
c^\dagger_{0,i}c_{1,j} & = & \frac{1}{\sqrt{2}} \left(c^\dagger_{1,i}c_{+,j}-c^\dagger_{1,i}c_{-,j}\right).
\end{eqnarray}
By simply combining these diagrams, and introducing ``twists'' in the chains, we can construct the devices presented in the body of this work. Of course, this is just a limited number of the countless possibilities.

\section*{APPENDIX B: REMOVING A SITE WITH A GATE POTENTIAL}

In this Appendix we give a closer look at the link term connecting the two chains:
\begin{equation}
H_\mathrm{link} = (-t_q+\frac{V_g}{2}) \left(c_{0,0}^\dagger c_{1,0} + \mathrm{h.c.}\right) + \frac{V_g}{2} \left(n_{0,0}+n_{1,0}\right).
\end{equation}
After the considerations introduced in the previous Appendix, it is easy to see that this Hamiltonian can be transformed into:
\begin{eqnarray}
H_\mathrm{link} & = & (-t_q+\frac{V_g}{2}) \left(n_{+,0}-n_{-,0}\right) + \frac{V_g}{2} \left(n_{+,0}+n_{+,0}\right) \nonumber \\
& = & -t_q \left(n_{+,0}-n_{-,0}\right) + V_g n_{+,0}.
\end{eqnarray}
When $t_q=0$, this reduces to:
\begin{equation}
H_\mathrm{link} = V_g n_{+,0}.
\end{equation}
For large $|V_g|$, this term will force the ``+'' orbital to be either fully occupied, or empty, with the consequent suppression of charge fluctuations. Therefore, all the transport will occur through the ``-'' orbital, meaning that the total system will behave as though it had one less site.

\section*{APPENDIX C: SYMMETRIES IN THE CONDUCTANCE}

\begin{figure}
\epsfig{file=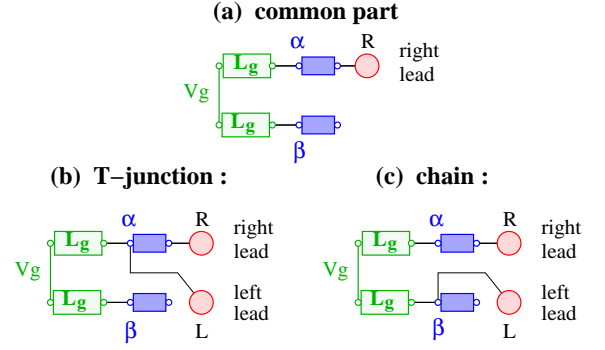,width=8.5cm}
\caption{Rearrangement or the different components of the circuit: (a) the common block for both systems. Connections for the T-junction (b) and for the chain (a) systems.}
\label{appendix1}
\end{figure}

In order to understand why the conductance is the same for the T-junction and the single chain, {\it i. e.} does not depend on the flux for $N=4k$, and $N=4k+2$, $L_g$ odd, we need to calculate the propagator $G_{L R}$ appearing in the expression for the conductance. We first realize that, independently of the connection, both systems have a common part shown in Fig.~\ref{appendix1}(a). Therefore, we calculate the partial propagators for this common segment, $\tilde{g}_{\alpha R}$ and $\tilde{g}_{\beta R}$, and we establish the connection to the second lead using a Dyson equation. 

For the T-junction, the left lead must be connected to the site labeled as $\alpha$, while for the the single chain, the connection must be established through site $\beta$. This is shown in Figures ~\ref{appendix1}(b) and~(c).

The equations of motion for the propagator $G_{L R}$ for the T-junction case can be written as,
\begin{equation}
\begin{cases}
 G_{L R}^{junction} = g_{LL} ~t_{L\alpha} ~G_{\alpha R}, \\
 G_{\alpha R} = \tilde{g}_{\alpha R} + \tilde{g}_{\alpha \alpha} ~t_{\alpha L} ~G_{L R}^{junction},
\end{cases}
\end{equation}
while for the chain we have,
\begin{equation}
\begin{cases}
 G_{L R}^{chain} = g_{LL} ~t_{L\beta} ~G_{\beta R}, \\
 G_{\beta R} = \tilde{g}_{\beta  R} + \tilde{g}_{\beta  \alpha} ~t_{\beta L} ~G_{L R}^{chain},
\end{cases}
\end{equation}
where $t_{L \alpha}=t_{L \beta}=t$, and we had preserved the subindex to distinguish the different cases. The bare Green functions $g_{ij}$ are the propagators before reestablishing the connection between sites $\alpha$ and $R$. The Green's functions $\tilde{g}_{ij}$ denote the propagators after reconnecting sites $\alpha$ and $R$ as shown in Fig.~\ref{appendix1}(a).

Solving these systems of equations, we obtain for the first,
\begin{equation}
G_{L R}^{junction} = \frac{g_{\alpha \alpha}}{1-g_{LL} ~t^2 ~\tilde{g}_{\alpha \alpha}} ~\left\{\frac{g_{RR} ~t^2 g_{LL}}{1-g_{RR} ~t^2 ~g_{\alpha \alpha}}\right\},
\label{sinF}
\end{equation}
and from the second set of equations,
\begin{equation}
G_{L R}^{chain} = \frac{g_{\beta  \alpha}}{1-g_{LL} ~t^2 ~\tilde{g}_{\beta  \beta}}  ~\left\{\frac{g_{RR} ~t^2 g_{LL}}{1-g_{RR} ~t^2 ~g_{\alpha \alpha}}\right\}
\label{conF}.
\end{equation}
where,
\begin{eqnarray}
\tilde{g}_{\alpha \alpha} &=&  g_{\alpha \alpha} + g_{\alpha \alpha} ~t ~\tilde{g}_{RR} ~t ~g_{\alpha \alpha} \label{partialsin}\\ 
\tilde{g}_{\beta  \beta } &=&  g_{\beta  \beta } + g_{\beta  \alpha} ~t ~\tilde{g}_{RR} ~t ~g_{\alpha \beta}  \label{partialcon}.
\end{eqnarray}

Comparing Eqs.~(\ref{sinF}) and~(\ref{conF}), we can observe that the term between brackets is common to both cases. The prefactors will determine when these two Green's functions are equal.
Note that by symmetry $g_{\alpha \alpha}=g_{\beta \beta}$. The prefactors in Eqs.~(\ref{sinF}) and~(\ref{conF}) as well as $\tilde{g}_{\alpha \alpha}$ and $\tilde{g}_{\beta \beta}$ from Eqs.~(\ref{partialsin}) and~(\ref{partialcon}) will be equal if,
\begin{equation}
g_{\alpha \alpha} = g_{\beta \alpha}.
\label{condition}
\end{equation}
Notice that, since we are calculating the zero bias conductance, this condition must be proved just for $\omega = 0$. 

%%%%%%%%%%%%%%%%%%%%%%%%%%%%%%%%%%%%%%%%%%%%%%%%%%%%%%%%%%%%%%%%%%%%55555

In order to demonstrate that the condition (\ref{condition}) is satisfied, we need to calculate the bare Green's functions for a single tight-binding segment of length $n=2L_g+(N-2)$.
For simplicity we will restrict this treatment to rings with $N=4$ sites, $n=2L_g+2$, but the demonstration can be extended to arbitrary ring sizes.
The Green's function is the inverse of the matrix $(H-Iz)$ where $H$ is the Hamiltonian of the segment. For the non-interacting case studied here, $H$ is a finite $n\times n$ symmetric tridiagonal matrix: $H_{ij} = a_i$ for $i=j$, and $H_{ij} = b_{i}$ for $i = j \pm 1$. There is a simple and elegant formula to calculate the inverse of such matrix \cite{Usmani1994}:
\begin{eqnarray}
g_{ij} = (-1)^{i+j}b_i \cdots b_{j-1}\theta_{i-1}\phi_{j+1}/\theta_n \,\,\,\,\, \mathrm{for}\,\, i \le j, \nonumber \\
g_{ij} = (-1)^{i+j}b_j \cdots b_{i-1}\theta_{j-1}\phi_{i+1}/\theta_n \,\,\,\,\, \mathrm{for}\,\, i > j, \nonumber
\label{gij}
\end{eqnarray}
where the $\theta_i$'s verify the recurrence relation
\[
\theta _i = a_i \theta_{i-1}-b_{i-1}^2\theta_{i-2}, \,\,\,\,\, \mathrm{for } \,\, i=2,\cdots ,n,
\]
with $\theta_0=1$ and $\theta_1=a_1$, and the $\phi_i$'s obey the recurrence
\[
\phi_i = a_i \phi_{i+1} - b_i^2\phi_{i+2}, \,\,\,\,\, \mathrm{for } \,\, i=n-1,\cdots ,1,
\]
with $\phi_{n+1} = 1$ and $\phi_n = a_n$. Notice that $\theta_n = \mathrm{det}(H-Iz)$. 

We are interested in the Green's function at the Fermi level, $z=0$ (we have ignored any imaginary contributions for simplicity). Moreover, all the diagonal elements are $a_i=0$ except $a_{n/2}=a_{n/2+1}=V_g/2$. Similarly, all the off-diagonal elements are $b_i=-1$, with the exception of $b_{n/2}=V_g/2$.
Using these expressions, it is easy to verify that $\theta_n=\mathrm{det}(H) = 0$ for $n=4k+2$, and $\theta_n=\mathrm{det}(H) = 1$ for $n=4k$. Moreover, for $n=4k$,
\[
g_{n1} = (-1)^{1+n}b_1 \cdots b_{n-1} = -V_g/2, 
\]
and
\[
g_{11} = (-1)^{n/2+1}b_{n-2}^2b_{n-4}^2\cdots b_{n/2}^2a_{n/2}b_{n/2-2}^2\cdots b_4^2b_2^2 = -V_g/2.
\]

Therefore, we can see that the recursions lead to the same numerical result for $g_{11}$ and $g_{n1}$ as long as the length of the chain obeys $n=4k=2L_g+2$, or $L_g$ odd. The important condition to get this equivalence is the presence of the symmetric chemical potentials $V_g$ in $H_{\mathrm{link}}$. Notice that these observations do not imply a symmetry relating the Hamiltonians of the chain and the T-junction, but a symmetry in the Green's functions at the Fermi level and, consequently, their equilibrium transport properties.

%%%%%%%%%%%%%%%%%%%%%%%%%%
%\bibliography{transistor}

\end{document}